\documentclass[12pt,letterpaper]{iopart}
\usepackage{iopams}  

\usepackage{graphicx}
\usepackage{cite}

\begin{document}

\title{Correlation function of weakly interacting bosons in a disordered lattice}

\author{B Deissler$^1$, E Lucioni$^1$, M Modugno$^{2,3}$, G Roati$^1$, L Tanzi$^1$, M Zaccanti$^{1,4,5}$, M Inguscio$^1$ and G Modugno$^1$}

\address{$^1$ LENS and Dipartimento di Fisica, Universit\`a di Firenze,
and INO-CNR, 50019 Sesto Fiorentino, Italy}
\address{$^2$ Department of Theoretical Physics and History of Science, UPV-EHU, 48080 Bilbao, Spain} 
\address{$^3$ IKERBASQUE, Basque Foundation for Science, 48011 Bilbao, Spain}
\address{$^4$ Institut f\"ur Quantenoptik und Quanteninformation,
 \"Osterreichische Akademie der Wissenschaften, 6020 Innsbruck, Austria}
\address{$^5$ Institut f\"ur Experimentalphysik, Universit\"at Innsbruck, 6020 Innsbruck, Austria}
\ead{deissler@lens.unifi.it}
\begin{abstract}
One of the most important issues in disordered systems is the interplay of the disorder and repulsive interactions. Several recent experimental advances on this topic have been made with ultracold atoms, in particular the observation of Anderson localization, and the realization of the disordered Bose-Hubbard model. There are however still questions as to how to differentiate the complex insulating phases resulting from this interplay, and how to measure the size of the superfluid fragments that these phases entail. It has been suggested that the correlation function of such a system can give new insights, but so far little experimental investigation has been performed. Here, we show the first experimental analysis of the correlation function for a weakly interacting, bosonic system in a quasiperiodic lattice. We observe an increase in the correlation length as well as a change in shape of the correlation function in the delocalization crossover from Anderson glass to coherent, extended state. In between, the experiment indicates the formation of progressively larger coherent fragments, consistent with a fragmented BEC, or Bose glass.
\end{abstract}

\pacs{34.50.-s, 37.10.Jk, 61.44.Fw, 67.85.Hj}
\maketitle

\section{Introduction}
The interplay of disorder and interactions lays at the heart of the behaviour of many physical systems, including superfluid helium in porous media\cite{Reppy 92}, granular and thin-film superconductors\cite{Dubi07,Beloborodov07,Goldman98,Phillips03}, and light propagating in disordered media\cite{Pertsch04,Schwartz07,Lahini08}. A central aspect for ultracold bosonic systems is the competition between disorder, which tends to localize particles, and weak repulsive interactions, which instead have a delocalizing effect. Wheareas disorder tends to localize non-interacting particles giving rise to Anderson localization\cite{Anderson58}, weak repulsive interactions can counteract this localization in order to minimize the energy. Eventually, interactions can screen the disorder\cite{Lee90} and bring the system towards a coherent, extended ground state, i.e.\ a Bose-Einstein condensate (BEC).

Systematic experimental studies of this interplay are difficult in condensed matter systems, since interactions are strong but difficult to control\cite{Reppy92}, while on the other hand in photonic systems only non-linearities corresponding to attractive interactions\cite{Schwartz07,Lahini08} have been explored in experiments. Instead, ultracold atoms in disordered optical potentials are a promising system for such investigations\cite{Fallani08,Sanchez10} due to their unprecedented control over the disorder strength and interactions. In fact, they have already enabled the observation of Anderson localization for bosons in the regime of negligible interactions\cite{Billy08,Roati08}, and recent experiments have investigated the effect of interactions on the localization properties, both in the weakly interacting\cite{Chen08,Deissler10} and strongly correlated\cite{Fallani07,White09,Pasienski10} regimes. 

Many theoretical predictions have been made about the properties of the complex phases appearing in these systems\cite{Giamarchi88,Fisher88,Fisher89,Scalettar91,Damski03,Schulte05,Lugan07,Roux08,Falco09,Falco09a,Deng09,Pollet09,Gurarie09,Cai10}. In particular, various methods to characterize these phases experimentally have been proposed, including measurements of transport properties\cite{Sanchez08,Adhikari09}, condensate and/or superfluid fractions\cite{Damski03,Schulte05,Roux08,Schmitt09}, excitation spectrum\cite{Schmitt09}, overlap function\cite{Morrison08}, and compressibility\cite{Roscilde09,Delande09,Shrestha10}. Recent interest has been in the correlation properties of disordered, interacting bosonic systems\cite{Fontanesi09,Fontanesi10,Cetoli10,Radic10}, in order to differentiate insulating phases such as the Bose glass from the superfluid regime. 

Here, we expand upon and extend our previous experimental work on bosons in a bichromatic optical lattice\cite{Deissler10}, with an emphasis on the correlation properties of our system. We measure the localization properties, spatial correlations and coherence properties of neighbouring states as a function of the interaction energy and study the delocalization crossover in terms of these observables. In addition, we study in detail the long-range decay of the correlation function of our system. Our data provide evidence of a change of decay behaviour at the crossover between insulating and superfluid phases, in agreement with theoretical predictions.

The paper is organized as follows: In \sref{sec:lattice}, we give an introduction to the physics of a quasiperiodic lattice, and describe the expected effects of repulsive interactions on bosonic atoms therein. In \sref{sec:experiment}, we detail the experimental scheme, before describing the image analysis methods employed and extracted observables in \sref{sec:analysis}. After showing the experimental results and comparing to theoretical predictions in \sref{sec:results}, we summarize and give an outlook in \sref{sec:conclusions}.

\section{Disordered phases and quasiperiodic optical lattices}\label{sec:lattice}

\begin{figure}
\begin{center}
\includegraphics[width=0.8\textwidth]{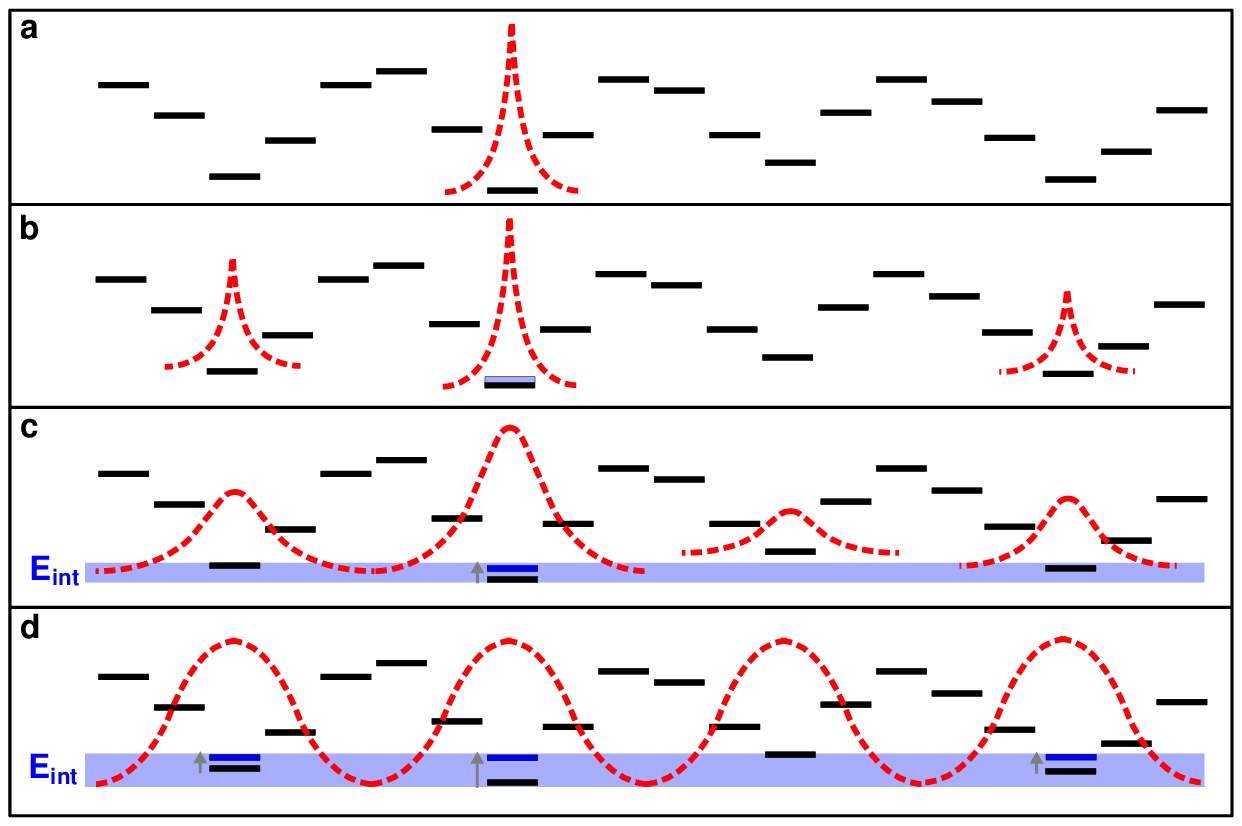}
\caption{Schematic of the interaction-induced delocalization in a quasiperiodic lattice. In the non-interacting case, the eigenstates are exponentially localized for sufficiently large disorder, and the absolute lowest energy state is populated (a). Introducing very weak repulsive interactions, several of the lowest energy states are populated (Anderson glass, b). The energies of different states can become degenerate due to repulsive interactions and their shape might be modified, giving rise to the formation of locally coherent fragments (fragmented BEC, c), though global phase coherence is not restored until  the entire system forms a coherent, extended state (BEC, d) at large interaction strengths.}
\label{fig:Cartoon}
\end{center}
\end{figure}

Interactions have a profound effect on disordered systems. A schematic of the effect of weak repulsive interactions can be seen in \fref{fig:Cartoon} for the specific case of bosons in a quasiperiodic lattice, as considered in this paper. Non-interacting bosons condense into the absolute lowest energy state of the disordered potential (\fref{fig:Cartoon}a). The defining characteristic of this Anderson localized state is its exponential shape. Adding repulsive interactions is expected to have a delocalizing effect. This can be understood in terms of a screening argument\cite{Lee90}. Repulsive interactions serve to smooth over the disordering potential in the occupied sites, providing a flatter energetic landscape on which more extended states can form. For very weak interactions, several low energy eigenstates of the non-interacting system can become populated (\fref{fig:Cartoon}b). This regime, in which several exponentially localized states coexist without phase coherence, is often identified with an Anderson glass\cite{Scalettar91,Damski03}, or Lifshitz glass\cite{Lugan07}. At larger interaction energies, an increasing number of sites is occupied, including neighbouring wells. When these states overlap, locally coherent fragments are expected to form (\fref{fig:Cartoon}c). In this case, global phase coherence would not yet be restored, and the local shape of the states might be modified. Some authors have called this regime a `fragmented BEC'\cite{Lugan07} or Bose glass\cite{Fisher88,Roux08,Fontanesi09}. The number of independent fragments should decrease for larger interaction energies, until finally, for sufficiently large interaction strengths a single, extended phase-coherent state is formed, that is, a macroscopic BEC (\fref{fig:Cartoon}d). The centre of the crossover from localized to extended, coherent states is expected to occur when the interaction energy is comparable to the standard deviation of disorder energy.

\begin{figure}
\begin{center}
\includegraphics[width=\textwidth]{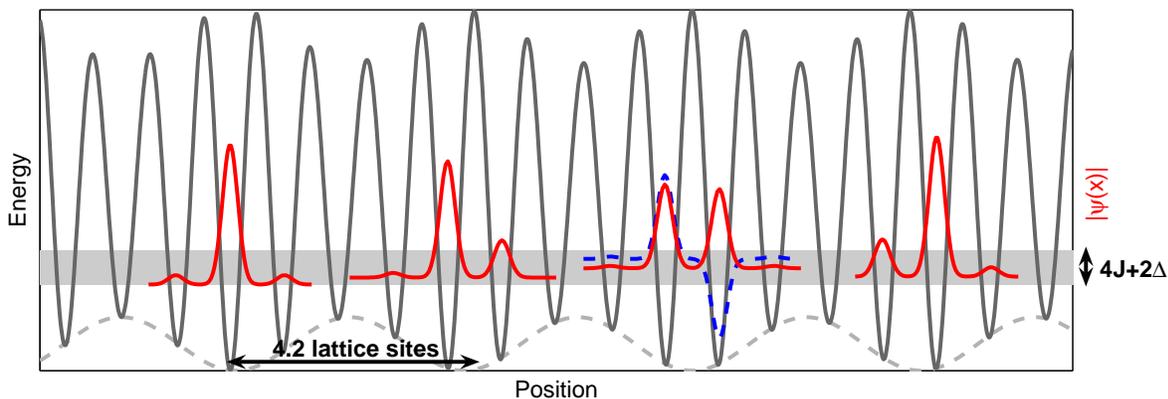}
\caption{Quasi-periodic potential. The quasi-periodic potential realized in the experiment for lattice incommensurability $\beta = 1.238\ldots$ and $\Delta/J = 6.2$. The gray stripe shows the energy of the first band of the combined lattice, with width $2\Delta+4J$. The quasi-periodic lattice is characterized by potential wells approximately every $1/(\beta-1) \approx 4.2$~lattice sites, which arise from the beating of the two lattices (grey dashed line). In red, the lowest energy eigenstates of the lattice potential are shown. Where two lattice sites are nearly equal in energy in a characteristic well, the potential looks locally like a double well, and there exist symmetric (red) and anti-symmetric (blue dashed) eigenstates, with an energy splitting of $2J$.}
\label{fig:bichromatic}
\end{center}
\end{figure}

Our system uses a particular kind of disorder, namely a quasiperiodic potential. This consists of two overlapping lattices with incommensurate wavelengths. The resulting potential can be seen as a strong primary lattice of periodicity $d = \pi/k_1$ which is perturbed by a weaker secondary lattice of periodicity $\pi/k_2$ ($k = 2\pi/\lambda$, where $\lambda$ is the wavelength of the light generating the lattice). The lattice potential can then be written as
\begin{equation}\label{eq:potential}
    V(x)=s_{1}E_{R} \sin^{2}(k_{1}x)+s_{2}\beta^2 E_{R} \sin^{2}(k_{2}x+\phi),
\end{equation}
where $E_{R}=\hbar^{2}k_1^{2}/(2M)$ is the recoil energy for the primary lattice ($M$ is the atomic mass), $\beta = k_2/k_1$, and $s_{i}$ are the heights of the lattices in units of their recoil energies. The lattice spacing of such a potential is to good approximation given by that of the primary lattice $d = \lambda_1/2$\cite{Guarrera07}.
The essential features of such a potential are visible in \fref{fig:bichromatic}. The potential energy minima of the primary lattice are modulated by the second one, giving rise to characteristic wells separated on average by $d/(\beta-1)$. The additional structure given by these characteristic wells can be employed in the analysis of the experimental data, as described in \sref{sec:analysis}.

For non-interacting atoms, the full Hamiltonian can be mapped onto that of the Harper\cite{Harper55} or Aubry-Andr\'e model\cite{Aubry80,Aulbach04,Boers07,Modugno09},
\begin{equation}
H = -J \sum_j \left( c_{j+1}^* c_j + c_j^* c_{j+1} \right) + \Delta \sum_j \cos(2\pi\beta j + \phi) |c_j|^2,
\end{equation}
where $j$ is a label for the lattice sites, and $c_j$ give the amplitude of the Wannier state centered at site $j$. In a tight-binding model, the tunneling energy is that of the primary lattice, and can be calculated in terms of the experimental parameters as\cite{Gerbier05}
\begin{equation}
J = 1.43 s_1^{0.98} \exp\left\{-2.07\sqrt{s_1}\right\} E_{R}.
\end{equation}
The disorder energy can be obtained from a numerical calculation as\cite{Modugno09}
\begin{equation}
\Delta = 0.5 s_2 \beta^2 \left[ \exp \left(-2.18/s_1^{0.6} \right) \right] E_{R}.
\end{equation}
The Aubry-Andr\'e model displays a transition at $\Delta/J = 2$ from extended to localized eigenstates. In an experimental realization with sufficiently large primary lattice, the localized regime is characterized by the absence of mobility edges, as well as exponentially localized eigenstates with the same localization length. It should be noted that this differs both from the case of a randomly disordered system, for which any non-zero amount of disorder is sufficient to localize the system in one dimension\cite{Kramer93}, and from the case of a speckle potential, for which effective mobility edges exist due to the correlated disorder\cite{Lugan09}.

\begin{figure}
\begin{center}
\includegraphics[width=0.7\textwidth]{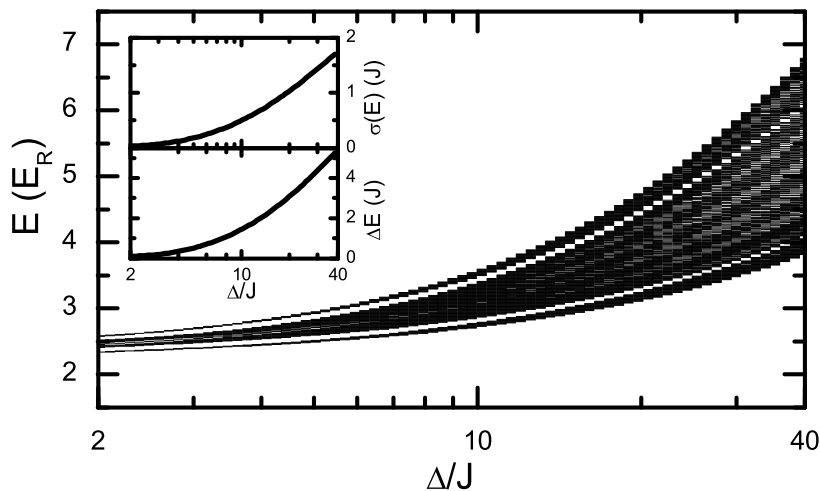}
\caption{Energies of the eigenstates of the quasiperiodic potential in absence external confinement as a function of $\Delta/J$ for $s_1 = 6.6$. Inset top: standard deviation of energies in the first ``miniband''; inset bottom: extension of first ``miniband''.}
\label{fig:Espectrum}
\end{center}
\end{figure}

The spectrum of such a quasi-periodic potential can easily be calculated and is shown in \fref{fig:Espectrum} for various values of the disorder strength $\Delta/J$ above the localization transition, neglecting any external confining potential. A first striking feature is the appearance of minigaps in the spectrum, the lowest of which has approximately the same width for all values of $\Delta/J$. The appearance of minigaps can be understood from the distribution of lattice sites in the characteristic potential wells (see \fref{fig:bichromatic}). The absolute lowest energy eigenstates are those for which a lattice site coincides with the minimum of a potential well. In contrast, when two neighbouring lattice sites are nearly symmetric in potential energy in a well, the potential appears locally like a double well, for which the two lowest-lying eigenstates have a separation in energy of $2J$. In fact, the width of the lowest minigap is approximately $2J$ throughout the range of $\Delta/J$ shown. The lowest ``miniband'' of energies corresponds to the lowest energy eigenstates localized in the potential wells $4.2d$ apart. Since in the experiment, only the states in the first ``miniband'' are populated, we restrict our analysis to these energies and show their standard deviation $\sigma(E)$ and the extension $\Delta E$ of this band. 
The effect of a confining potential on the spectrum has been analysed previously in ref.~\cite{Modugno09}.

\section{Experimental methods}\label{sec:experiment}
In the experiment, a degenerate Bose gas of $^{39}$K is employed in a quasiperiodic optical lattice. The production of a BEC of $^{39}$K has been described in detail previously\cite{Roati07}. A broad Feshbach resonance allows a tuning of the interactions, and even a nearly complete cancellation\cite{DErrico07}. In our case, a BEC of 40 000 atoms at a scattering length of $250 a_0$ is initially prepared in a crossed dipole trap. The condensate is loaded into the quasi-periodic potential while the optical trap is decompressed in about 250 ms to reduce the harmonic confinement, and a gravity-compensating magnetic field gradient is added. At the same time, the scattering length $a$ is changed by means of the broad Feshbach resonance to values ranging from $a\leq 0.1 a_0$ to about $a =300 a_0$. At the end of this procedure, the lattice lasers give a harmonic confinement of $\omega_\perp = 2\pi \times 50$~Hz in the radial direction. In the vertical (axial) direction, a weak confinement of 5~Hz is given by a weak optical trap as well as by a curvature from the gravity-compensating magnetic field. The primary lattice is generated by a Nd:YAG laser with a wavelength $\lambda_1 = 1064.4$~nm and a strength $s_1 = 6.6$, which is well within the tight binding regime. The secondary lattice is generated by a Ti:Sapphire laser of wavelength $\lambda_2 = 859.6$~nm and variable strength up to $s_2 = 1.2$. For these experimental parameters, the separation of neighbouring states is given on average by $d/(\beta-1) \approx 4.2$ lattice sites.

We estimate that around 30 lattice sites, corresponding to about 7 adjacent localized states, are populated during the loading of the lattice. We then define a mean interaction energy per particle $E_{\mathrm{int}} = gN/7 \int |\varphi(\mathbf{r})|^4 \, \rmd^3\mathbf{r}$, where $g = 4\pi\hbar^2a/m$ and $\varphi(\mathbf{r})$ is a Gaussian approximation to the on-site Wannier function. We include coupling into the radial directions of our system, with the consequence that the interaction energy is non-linear in the scattering length. Though this definition of the energy is strictly valid only in the localized regime, comparison with a numerical simulation of our experimental procedure has shown that it is a good approximation for all values of the scattering length up to an error of 30\%.

\begin{figure}
\begin{center}
\includegraphics[width=0.7\textwidth]{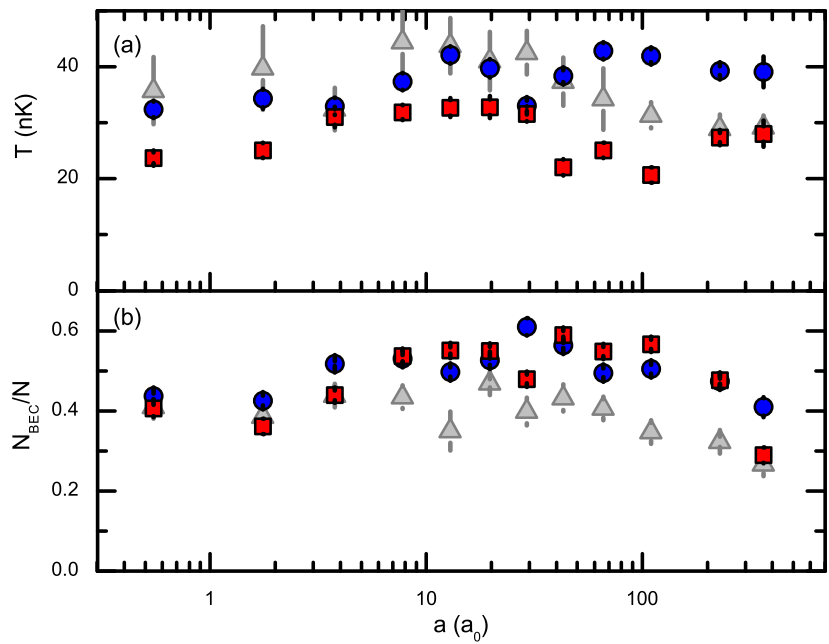}
\caption{Radial temperature (a) and condensed fraction (b) as extracted from a fit to the radial profile. The blue circles are for $\Delta/J = 6.2$, the red squares for $\Delta/J = 9.5$. The grey triangles are for a single lattice ($s_1 = 5.7$). The error bars denote the standard error of the mean.}
\label{fig:Radial}
\end{center}
\end{figure}

The loading process is adiabatic for most of the parameter range explored until $E_{\mathrm{int}}$ becomes sufficiently low for the system to enter the fully localized regime\cite{Edwards08}. Here, several independent low-lying excited states are populated even when it would be energetically favourable to populate just the ground state. A loss of adiabaticity can be seen experimentally as a transfer of energy into the radial direction. We measure a radial temperature and condensed fraction by fitting the radial profiles extracted from the absorption images with a two-component fitting function. In our previous work\cite{Deissler10}, a radial heating was seen to occur in a region of large disorder and weak interactions. In this work, the disorder strength is smaller and as a consequence the radial temperature as well as the condensate fraction are approximately constant throughout the parameter range explored (\fref{fig:Radial}).

\section{Momentum distribution, correlation function and phase fluctuations}\label{sec:analysis}
The system can be characterized by analyzing its momentum distribution and derived Fourier transforms. These techniques are used to extract information both about the local shape of the wavefunction, spatial correlations, and the coherence properties of neighbouring states. An image of the momentum distribution is taken by absorption imaging with a CCD camera after 46.5~ms ballistic expansion. This time is sufficiently long in order to be in the ``far-field'' limit\cite{Gerbier08}. At the time of release, the scattering length is set to below $1a_0$ in less than one ms and kept there until the Feshbach magnetic field is switched off 10~ms before taking the image -- at this point, the system has expanded a sufficient amount to neglect the effect of interactions.
After such a long free expansion without interactions, the image of the atoms that is acquired is approximately the in-trap momentum distribution $\rho(k) = \langle \hat\Psi^\dagger(k) \hat\Psi(k) \rangle$ \cite{Bloch08}, where $\hat\Psi(k)$ is the Fourier transform of the bosonic field operator $\hat\Psi(x)$. In order to recover information about the in-trap wavefunction, we can therefore use an inverse Fourier transform.

\begin{figure}
\begin{center}
\includegraphics[width=\textwidth]{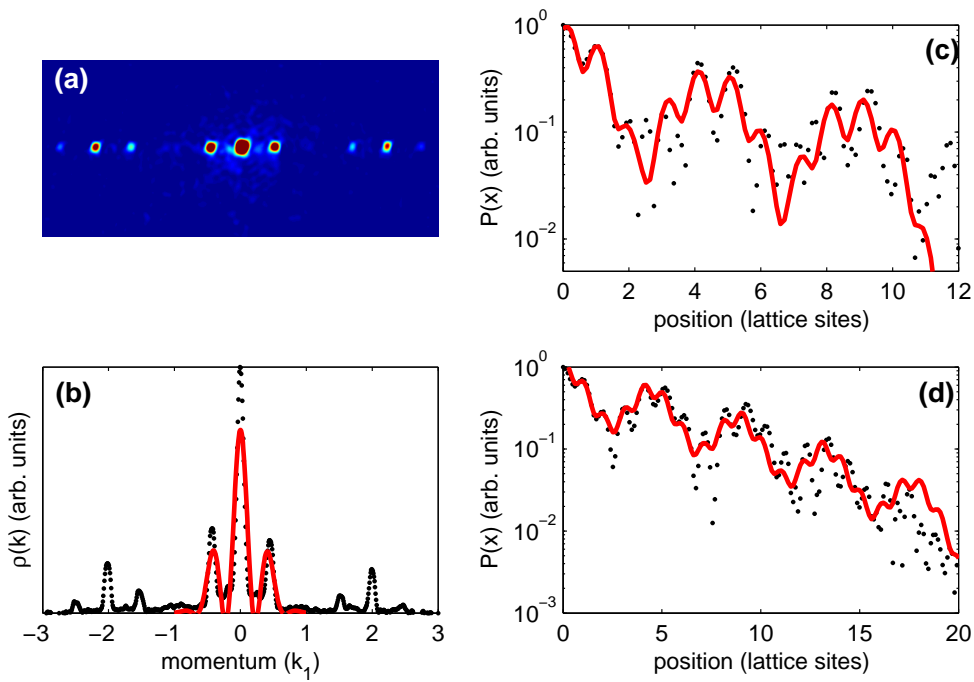}
\caption{Example of image analysis. After integration along the radial direction of the acquired absorption image (a), the profile of the momentum distribution is fit with a modulated Gaussian to recover the phase fluctuations (b). The Fourier transform of the square root of the profile can then be fit with three generalized exponential functions to extract the exponent and local length of the localized states (c). The correlation function $g(x)$ is given by the FT of the momentum distribution itself, and can be fit with two generalized exponentials, as in (c) in order to get $g(4.2d)$and $g(8.4d)$, or it can be fit with a generalized exponential decay up to 20 lattice sites (d).}
\label{fig:AnalysisFig}
\end{center}
\end{figure}

Due to the quasi-periodic nature of the employed lattice potential, we expect that for a sufficiently homogeneous system, the in-trap wavefunction can be decomposed into copies of the same state with real and non-negative envelope $\xi(x)$, spaced by $D = 4.2d$. The overall in-trap wavefunction can therefore be approximated as
\begin{equation}
 \psi(x) = \sum_j  a_j \xi(x-jD) e^{-i\phi_j}, 
\end{equation}
where $\phi_j$ is the local phase, and $\xi(x)$ can be taken as a generalized exponential function $\exp(-|x/L|^\alpha)$. In momentum space, the magnitude of the overall wavefunction can then be written as $\sqrt{\rho(k)} = |\xi(k)| \mathcal{S}(k)$, where
\begin{equation}
 \mathcal{S}(k) = \left| \sum_j a_j e^{-i(jkD + \phi_j)} \right| 
\end{equation}
is an interference term. For many envelope functions $\xi(x)$, such as the generalized exponentials with $0 < \alpha \leq 2$, the Fourier transform $\xi(k)$ itself is real and non-negative\cite{Giraud06}, so that the inverse Fourier transform of $\sqrt{\rho(k)}$ can be written as $\xi(x) \circ \mathcal{S}(x)$. This is simply the convolution of the envelope of a single state $\xi(x)$ with the Fourier transform of the interference term, $\mathcal{S}(x)$, which can be approximately described as a series of sharp peaks (approaching $\delta$-distributions) spaced by $D$, with a decreasing amplitude and phases that depend on the local phases $\phi_j$ and amplitudes $a_j$. 

The inverse Fourier transform of the square root of the momentum distribution $\rho(k)$ therefore allows an estimate of the average local shape of the (wave)function $\xi(x)$. The smallest cloud size observable with the imaging system employed is about 12~$\mu$m, therefore there is a finite resolution also in momentum space (about $k_1/35$). Due to this finite resolution, the Fourier transform has an envelope with a width of about 10 lattice sites. This means that we can only distinguish easily up to three neighbouring states. The averaged wavefunction is analyzed by fitting to the sum of three generalized exponential functions modulated by the primary lattice
\begin{equation}
f(x) = \left[ \sum_{j=0}^2 A_j \exp\left(-\left|\frac{x-jD}{L_s}\right|^\alpha\right) \right] \cdot \left[ 1+B\cos(k_1 x + \delta) \right],
\label{eq:expfit}
\end{equation}
see \fref{fig:AnalysisFig} for examples. From such a fit, the exponent $\alpha$ and the local extension of the states $L_s$ can be extracted.

On the other hand, the inverse Fourier transform of the momentum distribution itself can be employed to find the correlation properties of neighbouring states. 
The momentum distribution can be related to the first order correlation function $G(x',x+x') = \langle \hat \Psi^\dagger(x') \hat \Psi(x+x') \rangle$. We analyze the spatially averaged correlation function, which, using the Wiener-Khinchin theorem, can be expressed as
\begin{equation}
g(x) = \int G(x',x+x') \, \rmd x' 
	 = \int \frac{\rmd k}{2\pi} \rho(k) e^{ikx}.
\end{equation}
Experimentally, $g(x)$ is recovered simply by taking the Fourier transform of the momentum distribution. 
We fit with the same generalized exponential of \eref{eq:expfit} and recover the spatially averaged correlation between states 4.2 (8.4) lattice sites apart as $A_2/A_1$ ($A_3/A_1$). Also here, the finite momentum resolution limits our analysis to three neighbouring sites, and it follows that $g(4.2d)$ ($g(8.4d)$) saturates at a value around 0.6 (0.3).

More information about the extent and decay of the spatially averaged correlation function can be gained by examining the Fourier transform of the momentum distribution at larger distances. While the detailed structure there is not resolvable, making the data there unsuited for the analysis described above, we can extract information about the general shape of $g(x)$. The data is fit with a function
\begin{equation}
\fl g(x) = \left[ \sum_{j=0}^4  \exp \left( - \left| \frac{jD}{L_g} \right|^\beta \right) \cdot
	\exp \left(- \left| \frac{x-jD}{L_s} \right|^\alpha \right) \right] 
\cdot	\left[ 1 + B\cos(k_1 x + \delta) \right],
\end{equation}
where $\beta$ is the correlation exponent and $L_g$ is the correlation length. This describes the sum of five generalized exponential functions spaced by $D$, with amplitudes determined by the shape of the correlation function. Note that this correlation function is called degree of coherence by Fontanesi \textit{et al}.\ in ref.~\cite{Fontanesi09}.

Finally, the effect of a fluctuating phase between neighbouring states is seen as a shift of the phase $\phi$ of the interference in the momentum distribution. We extract this phase by fitting the momentum distribution within the first Brillouin zone directly with a fitting function
\begin{equation}
A \exp\left(-\frac{(k-k_C)^2}{2w^2}\right) \cdot \left[ 1+B\cos\left(D(k-k_C) + \phi \right) \right],
\end{equation}
where $k_C$ is the center of the distribution, determined by fitting the average of all images of a given dataset.

\section{Observed disordered regimes}\label{sec:results}
\begin{figure}
\begin{center}
\includegraphics[width = \textwidth]{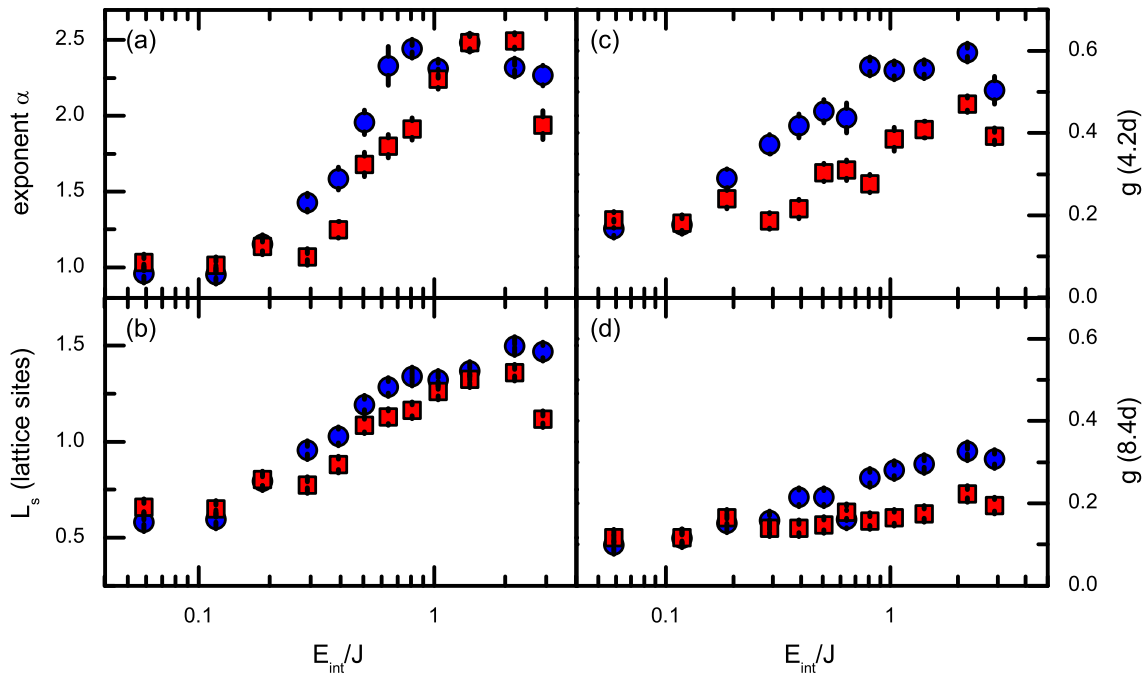}
\caption{Results of the Fourier analysis. The blue circles are for $\Delta/J = 6.2$, the red squares for $\Delta/J = 9.5$. The exponent $\alpha$ (a) and local length $L$ (b) are extracted from a fit of three generalized exponential states to the FT of the square root of the momentum distribution. The spatially averaged correlation $g(x)$ is extracted from the FT of the momentum distribution itself, and fits with three generalized exponential functions evalate $g(x)$ at $4.2d$ (c) and $8.4d$ (d). The error bars denote the standard error of the mean.}
\label{fig:FFTresults}
\end{center}
\end{figure}

We present measurements of the quantities described in the previous section for two cuts of the phase diagram shown in ref.~\cite{Deissler10}. This data was taken for $s_1 = 6.6$, corresponding to a value of the tunneling energy of $J/h = 200$~Hz, and for two values of the disorder strength, $\Delta = 6.2J$ and $\Delta = 9.5J$. Both lattice strengths were calibrated by Bragg diffraction\cite{Cronin09}, with an estimated error on $\Delta/J$ of around 15\%.

The results of the analysis of localization properties as well as correlation of nearest and next-nearest neighbouring states are shown in \fref{fig:FFTresults}, where we plot the local extension of the states $L_s$, the exponent $\alpha$, and the correlation function evaluated at $4.2d$ and $8.4d$, as a function of the interaction energy $E_{\mathrm{int}}$.
We find that for very small $E_{\mathrm{int}}$, the states are exponentially localized, since the exponent $\alpha\approx 1$, and the local length $L_s$ is small, consistent with the Anderson glass regime. Increasing $E_{\mathrm{int}}$, the local length increases and the exponent increases up to $\alpha > 2$. Repulsive interactions therefore delocalize the system as expected, or alternatively, the localization crossover is shifted to higher values of the disorder strength $\Delta/J$ when interactions are introduced into the system.
In the localized regime, the correlation is finite but small, due to the occupation of independent neighbouring localized states arising from the non-adiabatic loading. As $E_{\mathrm{int}}$ is increased, the correlation features a crossover towards larger values, signalling that coherence is progressively established locally over distances of first $4.2d$ and then $8.4d$. The position of this crossover is in good agreement with the prediction of the simple disorder screening argument, from which we expect the centre of the crossover to occur when $E_{\mathrm{int}}$ is comparable to the standard deviation of energies in the lowest miniband. For $\Delta/J = 6.2$ ($\Delta/J = 9.5$), this is given by $0.26J$ ($0.47J$).

\begin{figure}
\begin{center}
\includegraphics[width=0.7\textwidth]{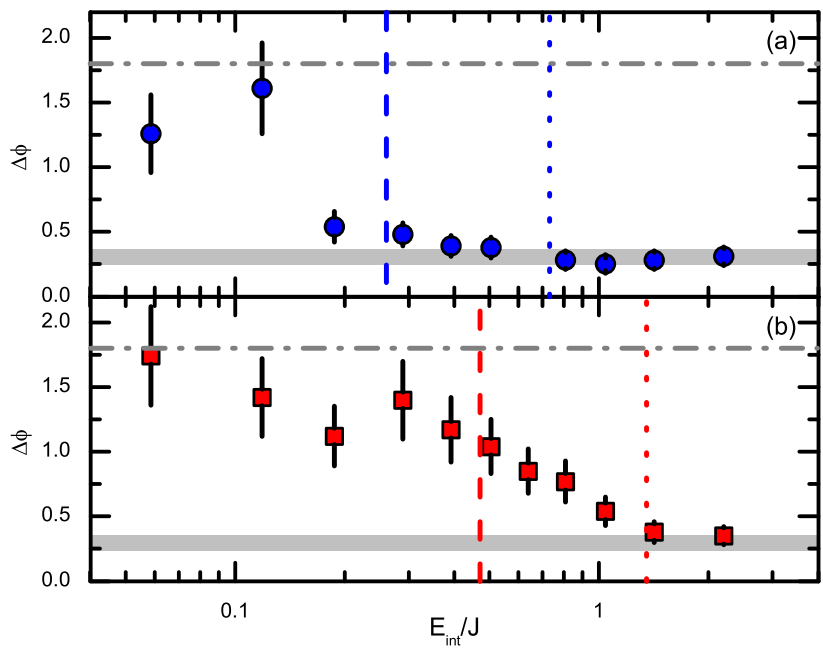}
\caption{Standard deviation of the phase measured by repeating the experiment up to 26 times for a given set of parameters, for $\Delta/J = 6.2$ (a) and $\Delta/J = 9.5$ (b). The error is estimated as $\Delta\phi/\sqrt{N}$, where $N$ is the number of images from which the phase was extracted. The grey shaded bar shows the phase fluctuations measured for an extended system below the localization threshold. The dash-dotted line gives the standard deviation for a purely random distribution. The dashed (dotted) lines gives the standard deviation (full extension) of energies in the lowest miniband.}
\label{fig:PhaseFluctuations}
\end{center}
\end{figure}

Information about the phase coherence of neighbouring states can be obtained by measuring the phase $\phi$ of the interference pattern in the momentum distribution for repeated runs of the experiment with the same parameters. If the states are not phase locked, $\phi$ changes almost randomly at each repetition of the experimental sequence. In \fref{fig:PhaseFluctuations} we show the standard deviation of $\phi$, estimated from a large number of repetitions of the experiment. We see a decrease of the phase fluctuations with increasing $E_{\mathrm{int}}$, that nevertheless remain relatively large in the crossover region where the correlation increases. The fluctuations finally drop to the background value only when $E_{\mathrm{int}}$ is comparable to the full width of the lowest miniband of the non-interacting spectrum (dotted lines in \fref{fig:PhaseFluctuations}). These observations confirm that in the localized regime the states are totally independent, which together with the localization properties (\fref{fig:FFTresults}) indicates that the system can indeed be described as an Anderson glass\cite{Scalettar91,Damski03}. The system crosses a large region of only partial coherence while becoming progressively less localized as $E_{\mathrm{int}}$ is increased. This is consistent with the formation of locally coherent fragments expected for a fragmented BEC. An analogous fragmentation behaviour was reported in ref.~\citen{Chen08}. Ultimately, the features of a single extended, coherent state are seen, i.e.\ a BEC.

\begin{figure}
\begin{center}
\includegraphics[width=0.7\textwidth]{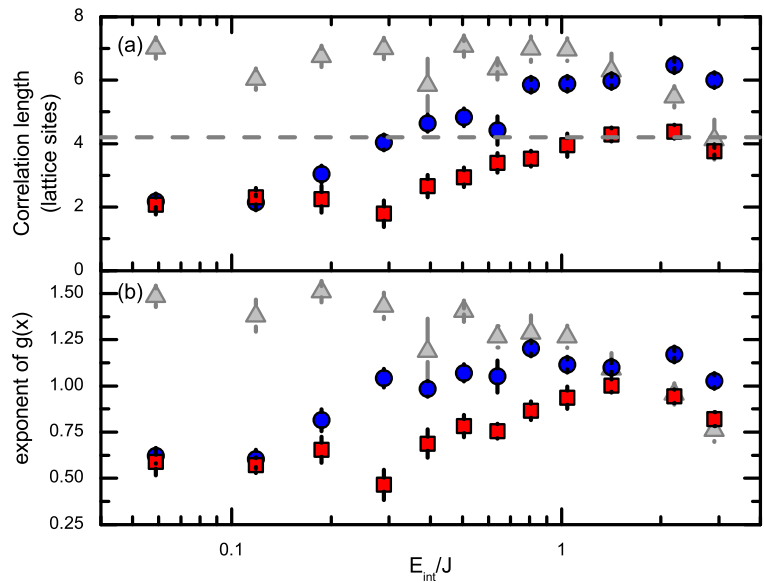}
\caption{Decay of the correlation function. (a) Correlation length, (b) exponent of the generalized exponential function. The blue circles are for $\Delta/J = 6.2$, the red squares for $\Delta/J = 9.5$. The grey triangles are for a single lattice ($s_1 = 5.7$), fit to a single generalized exponential function. The dashed line denotes the average separation of states, $4.2d$. The error bars denote the standard error of the mean.}
\label{fig:Correlations}
\end{center}
\end{figure}

Finally, we show the overall behaviour of the correlation function in \fref{fig:Correlations}. The correlation length $L_g$ increases at larger interaction energy to values larger than the mean separation of states ($4.2d$). It saturates at values around $6d$, consistent with the imaging resolution, for both quasiperiodic and single lattice potentials. The increase in correlation length shows that the average size of fragments found in the fragmented BEC regime increases with $E_{\mathrm{int}}$ until presumably only a single fragment describes the system.

The exponent $\beta$ of $g(x)$ is seen to increase from values of around $0.5$ to values slightly larger than $1$. While such an increase is qualitatively expected, the values of $\beta$ are not in agreement with expectations from theory. For a three-dimensional Bose gas at zero temperature, we expect a transition from exponential decay ($\beta = 1$) in the insulating regime\cite{Fisher89} to a shape of the correlation function given by the confining potential, $\beta \gtrsim 2$ in our case. In our analysis, any finite thermal component artificially reduces the exponent by increasing the values of the Fourier transform at small $x$ values. Indeed, we observe an exponent of $1.5$ or less even in the single lattice potential, for which the system is superfluid. In the quasiperiodic lattice, the exponent approaches that of the single lattice potential for large values of the interaction energy. The decrease of $L_g$ and $\beta$ at large values of $E_{\mathrm{int}}$ can presumably be explained by an imperfect removal of interaction energy from the system during the initial stage of expansion from the lattice. This would lead to a broader peak in the momentum distribution, and therefore a narrower shape at short distances in the correlation function.

In recent theoretical works on disordered bosonic systems, the change of shape of the correlation function in one dimension from exponential decay to algebraic decay was used to distinguish the Bose glass from the superfluid phase\cite{Fontanesi09,Cetoli10,Fontanesi10}. These theoretical investigations have the advantage of being able to consider large system sizes, where a jump in the first order correlation function $G(x_0,x)$ is an indication that fragments form, leading to an exponential decay of $g(x)$ in the Bose glass regime. In the experiment, the correlation function can only be recovered for smaller distances, due to the finite imaging resolution and system size. Fragments with sizes larger than approximately $2D$ cannot be distinguished from the superfluid. In this sense, the evolution of the shape of the correlation function can give information about the crossover from the Anderson glass to the fragmented BEC (where the correlation length starts to increase), but cannot quantify the crossover to the superfluid in our current system. Use of a higher resolution imaging system and eventually larger system sizes could enable the observation of the shape of the correlation function at larger distances. The crossover from fragmented BEC to superfluid could then be quantified.

\section{Conclusions and Outlook}\label{sec:conclusions}
In conclusion, we have characterized the entire delocalization crossover of a disordered bosonic system caused by weak repulsive interactions through study of the spatial localization, phase coherence, and correlation properties. In particular, we have shown the first experimental determination of the shape of the correlation function in such a system. We find three different regimes, in agreement with theoretical predictions. At vanishing interaction energy, the system can be described as an Anderson glass (or Lifshitz glass), with exponentially localized eigenstates without phase coherence between them. As $E_{\mathrm{int}}$ is increased, the local shape of the states changes, and coherence is gradually established, leading to an increase of the size of locally coherent fragments. This regime is consistent with a fragmented BEC, or Bose glass. Finally, for sufficiently large $E_{\mathrm{int}}$, the features of a single, extended coherent state are observed, and the system returns to a BEC. The position of the crossover is in good agreement with the predictions of a simple disorder screening argument for the lowest ``miniband''.

The techniques shown here are quite general and might be of use for further investigations of disordered systems. In particular, the analysis of the correlation function can also be used for experimental systems utilizing speckles\cite{Billy08,Chen08,Pasienski10}, for which the methods described here could easily distinguish superfluid and insulating phases. In the current experiment, the length scale over which the correlation function could be observed was primarily limited by the imaging resolution. More generally, the finite expansion time is expected to be a more important limitation\cite{Gerbier08}, especially when investigating larger sized systems. The required expansion time can easily be estimated by considering an in-trap wavepacket with a Gaussian width of $\Delta x_0$. In momentum space, this corresponds to a width $\Delta k = 1/\Delta x_0$. After a ballistic expansion for time $t_{\mathrm{exp}}$ after release of the wavepacket from the confining trap, this momentum component will move a distance $d_{\mathrm{exp}} = \hbar t_{\mathrm{exp}}/(M \Delta x_0)$. In order to see the features of the initial wavepacket, this distance must be larger than $\Delta x_0$, which implies that the expansion time must be larger than $M (\Delta x_0)^2/\hbar$ \footnote{This is $t_{\mathrm{FF}}$ in reference\cite{Gerbier08} for a coherent wavepacket.}. For our parameters, this is approximately 35~ms, less than the expansion time of 46.5~ms used. However, we must consider also our finite imaging resolution $\Delta x_{\mathrm{im}}$, which can be estimated to artificially increase the width of the initial cloud to $(\Delta x_{\mathrm{ex}})^2 = (\Delta x_0)^2 + (\Delta x_{\mathrm{im}})^2$. Given our resolution $\Delta x_{\mathrm{im}} \approx 12$~$\mu$m, this suggests a necessary expansion time of 120~ms, much longer than what is used in the experiment. However, a modest improvement of the imaging resolution to 5~$\mu$m would be sufficient to analyze the correlation function up to distances of 30 sites.

In the future, the Fourier analysis techniques might be used to explore the regime of strong correlations, $E_{\mathrm{int}} \gg J$, which can be reached by using a quasi-1D system with strong radial confinement. There, the Bose glass can be attributed to the \emph{cooperation} of disorder and interactions. There is however still debate on the exact shape of the phase diagram in this regime, particulary concerning the possibility of reentrant superfluidity\cite{Fisher89,Scalettar91,Damski03,Fallani07,Pollet09}. Furthermore, the analysis of the momentum distribution using Fourier techniques would also be of use in higher dimensional systems, where it might be possible to use phase retrieval algorithms to reconstruct the in-trap density distribution in detail\cite{Fienup78,Miao98}. 

\ack
We acknowledge stimulating discussions with L. Fontanesi and A. Cetoli. We thank C. D'Errico and M. Fattori for experimental contributions during the early stage of the experiment and discussions, J. Seman for experimental assistance and all the colleagues of the Quantum Gases group at LENS. This work has been supported by the EC (MEIF-CT-2004-009939 and Integrated Project AQUTE), by the ERC through the Starting Grant QUPOL and Advanced Grant DISQUA and by the ESF and INO-CNR through the EuroQUASAR and DQS EuroQUAM programs.

\end{document}